# The 'Big Three' of Scientific Information
## A comparative bibliometric review of Web of Science, Scopus, and OpenAlex

Daniel Torres-Salinas  & Wenceslao Arroyo-Machado

*EC3metrics, Universidad de Granada, Spain*

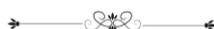

**Abstract**
The present comparative study examines the three main multidisciplinary bibliographic databases, *Web of Science Core Collection*, *Scopus*, and *OpenAlex*, with the aim of providing up-to-date evidence on coverage, metadata quality, and functional features to help inform strategic decisions in research assessment. The report is structured into two complementary methodological sections. First, it presents a systematic review of recent scholarly literature that investigates record volume, open-access coverage, linguistic diversity, reference coverage, and metadata quality; this is followed by an original bibliometric analysis of the 2015-2024 period that explores longitudinal distribution, document types, thematic profiles, linguistic differences, and overlap between databases. The text concludes with a ten-point executive summary and five recommendations.

**Keywords**
Literature review, Bibliometric analysis, Scientific information, Bibliographic databases, Web of Science, Scopus, OpenAlex, Coverage

**Author contribution**
Both authors contributed equally to all aspects of this work, including conceptualization, methodology, analysis, and writing.

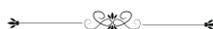

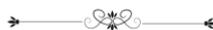

# Table of Contents

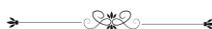

# INTRO

Over the past few years, the landscape of scientific information has undergone an unprecedented transformation that goes beyond the mere incorporation of new tools. This metamorphosis represents a structural shift in how scientific knowledge is produced, communicated, accessed, and evaluated. While just over a decade *ago* Web of Science and Scopus exercised virtually monolithic dominance in the retrieval and analysis of bibliographic information, the contemporary scenario is characterized by a multiplicity of information sources that offer different perspectives, broader coverage, and ultimately, new ways of retrieving and analyzing scientific activity (Torres-Salinas et al., 2023). Particularly, the last decade has witnessed the emergence of open databases, a phenomenon that strongly challenges the traditional commercial subscription model.

This process of change, to establish a specific milestone, began with Google Scholar (2004), which "democratized"[1] access to scientific literature through massive indexing of academic publications from diverse sources. This was complemented by the continued and sustained development of all types of open institutional repositories, driven by initiatives such as OpenDOAR[2], which facilitated self-archiving. More recently, OpenAlex, launched in 2022 by OurResearch as the successor to Microsoft Academic Graph, has positioned itself as a viable alternative through the integration of heterogeneous sources such as Crossref, PubMed, ORCID, and institutional repositories. Simultaneously, regional databases such as Dialnet and SciELO have consolidated themselves as infrastructures that provide visibility to Ibero-American scientific output in the humanities and social sciences, disciplines that have been historically marginalized. In parallel, alternative metrics (altmetrics) platforms such as Altmetric.com and PlumX have complemented this ecosystem by capturing the social impact of research beyond traditional citations.

This change is also closely linked to a transformation in research evaluation, initiated with the San Francisco Declaration on Research Assessment (DORA)[3] in 2012 and the Leiden Manifesto (Hicks et al., 2015), which questioned the use of the Impact Factor as a proxy metric for scientific quality. The Coalition for Advancing Research Assessment (CoARA)[4], promoted by the European University Association and the European Commission in January 2022, represents the institutionalization of this reform movement. In Spain, most research institutions have committed to these principles, now widely known, as have organizations such as ANECA, FECYT, and the State Research Agency (AEI). Therefore, the convergence between new open scientific information infrastructures and reforms in evaluation systems configures a scenario where the democratization of access is complemented by a more contextualized assessment of academic careers (Torres-Salinas, 2024). This is the landscape in which we operate and within which the present text is situated.

In this regard, the question is not which database to contract or which scientific information service to use, but rather when public institutions will take the final leap, without a safety net, toward open systems of publication and communication. As has been well demonstrated, if there is political will, it is possible to achieve a transition of the scientific community toward a world of open access (Arroyo-Machado & Torres-Salinas, 2025), free, above all, from the dramatic growth in article processing charges (APCs) (Haustein et al., 2024). This vision has been reinforced by the

---

[1] We employ the term "democratized" with deliberate ambiguity. Google Scholar does not originate open access to science; rather, it leverages the prior work of journals, repositories, and platforms funded with public resources. Its contribution lies in rendering visible that collective effort through a private infrastructure.
[2] https://opendoar.ac.uk/
[3] https://sfdora.org/read/read-the-declaration-espanol/
[4] https://www.coara.org/

Barcelona Declaration on Open Research Information (2024) [5], which commits prestigious academic and funding organizations to making openness in research information the new norm, recognizing that scientific access and evaluation should not be based on closed evidence managed by commercial infrastructures.

Although the long-term future is shaped by the institutional policies that our organizations are signing and by the emerging open universe of scientific information, the short- and medium-term scenario shows that we must continue walking along the usual paths. In this regard, commercial databases remain a defensible solution, as they are still of undeniable value to libraries, research communities, and R&D managers. Now more than ever, supporting our decisions and defending our curricula with reliable data and verified sources is essential, and we cannot always guarantee this with open sources. Therefore, to shed light on this landscape, we have drafted the present comparative analysis, at a basic level, of the main multidisciplinary bibliographic databases, the 'Big Three' of the moment: Web of Science, Scopus, and OpenAlex.

We can thus state that the main objective of the present study is to provide a comparison of the coverage, quality, and functional characteristics of the three databases. To this end, we have divided the report into two clearly differentiated parts:

**Part 1: Literature review**
In this section, we conduct a brief review of the literature. First, we introduce the reader to the main aspects of each of the databases, in the form of a technical sheet. Subsequently, we analyze relevant and recent academic works published on the three databases. In this section, we do not aim to be exhaustive, but rather to present information that may be pertinent to the context of the report. This analysis is complemented by a summary table of all aspects addressed.

**Part 2: Comparative bibliometric study**
To complement the previous review, a comparative bibliometric study is conducted with original and unpublished data from Web of Science, Scopus, and OpenAlex, using bibliographic data from the last decade (2015-2024) and focusing the analysis in depth on the most recent five-year period (2020-2024). This quantitative analysis essentially comprises a longitudinal study, differences by document typologies, thematic profiles, and differences by language.

The report follows a clear reading line: we introduce the databases, review the existing literature, conduct the quantitative analysis, which allows us to harmonize and verify figures offered in previous sections, and present conclusions. For those readers who wish to access the main conclusions directly, we have included an executive summary at the end of the document. Likewise, the report is accompanied by a methodological appendix where the details of the quantitative analysis conducted are specified. We hope to shed light on the subject.

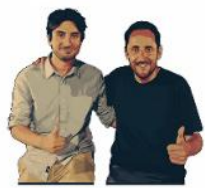

The Authors, in Granada, January 22, 2026

# PART 1

# Literature review

## 1.1. General and commercial overview

In this first part, we will analyze the main characteristics of the three databases that we will examine in the present report. The information we will use comes from professional and commercial websites, and with it we aim to offer an initial overview, especially directed at those who are most novice or less familiar with the databases. In all three cases, we will analyze basic information about: (1) business model, (2) quantitative coverage, (3) editorial criteria, and (4) analytical functionalities.

### 1.1.1. Web of Science of Core Collection - Clarivate Analytics

**Business model**

Web of Science Core Collection operates under a business model based on institutional subscription and access licenses, managed by Clarivate Analytics[6]. Its strategy focuses on offering different subscription tiers tailored[7] to the needs of academic institutions, with orientation toward universities, research centers, and public organizations that require reliable platforms for the evaluation of scientific activity. Clarivate maintains a recurring revenue model through long-term subscription agreements complemented by revenues from analytical services.

**Quantitative coverage**

Web of Science Core Collection indexes 22,704 journals[8]. The Science Citation Index Expanded (SCIE) covers more than 9,450 journals across 182 scientific disciplines, totaling 69 million records[9]; the Social Sciences Citation Index (SSCI) covers 3,541 journals across 58 disciplines and 11.8 million records[10]; the Arts & Humanities Citation Index (AHCI) indexes 1,808 journals across 25 categories and totals 5.6 million records[11]. Historical data dates back to 1900 in some disciplines and to 1945 in others. On the other hand, the Emerging Sources Citation Index (ESCI) contains 9,504 journals, 5.9 million records, and represents 252 categories, with the first year of indexing being 2005[12]. In total, the Core Collection, including the rest of the content such as books and proceedings, accounts for 97 million records[13].

Recently, in September 2025, the Research Commons[14] product has been incorporated into the platform, independently from the Core Collection, adding more than 32 million documents published in journals over the last 10 years. However, this is nothing other than the assimilation of OpenAlex, which we will discuss later, and it signals the duality of the company: on the one hand, it maintains a balanced strategy of curated growth within the Core Collection and breadth in coverage through the incorporation of complementary databases that, on the other hand, we already access freely through other means.

---

[6] https://clarivate.com/academia-government

[7] https://developer.clarivate.com/apis/

[8] https://clarivate.libguides.com/librarianresources/coverage

[9] https://clarivate.com/academia-government/scientific-and-academic-research/research-discovery-and-referencing/web-of-science/web-of-science-core-collection/science-citation-index-expanded/

[10] https://clarivate.com/academia-government/scientific-and-academic-research/research-discovery-and-referencing/web-of-science/web-of-science-core-collection/social-sciences-citation-index/

[11] https://clarivate.com/academia-government/scientific-and-academic-research/research-discovery-and-referencing/web-of-science/web-of-science-core-collection/arts-humanities-citation-index/

[12] https://clarivate.com/academia-government/scientific-and-academic-research/research-discovery-and-referencing/web-of-science/web-of-science-core-collection/emerging-sources-citation-index/

[13] https://clarivate.com/academia-government/scientific-and-academic-research/research-discovery-and-referencing/web-of-science/web-of-science-core-collection/

[14] https://clarivate.com/academia-government/blog/expanding-the-web-of-science-platform-with-research-commons/

**Editorial criteria**

The content selection in Web of Science Core Collection is governed by a rigorous editorial process led by Clarivate editors who operate independently, without affiliation to publishing houses or research institutions, thus avoiding potential conflicts of interest[15]. Regarding journal evaluation, it is based on 28 clearly defined criteria: 24 quality criteria designed to assess editorial rigor and good publication practices, and four impact criteria aimed at identifying the most influential journals using citations as the main indicator[16]. Perhaps this is one of the distinctive features of this database: the difficulty of being indexed, a criterion that has been relaxed in recent years. The database has incorporated the Emerging Sources Citation Index (ESCI), a "relaxed" index that allows the inclusion of journals of lower relevance that do not yet meet the impact levels required to enter the main indexes (SCIE, SSCI, AHCI).

It should be noted that the current evaluation process includes continuous monitoring of journal performance and possible exclusion due to misconduct or citation manipulation. In the June 2024 update, 17 journals were excluded from the Journal Citation Reports (JCR) and Web of Science Core Collection due to suspected citation manipulation and editorial "misconduct" (Attride & Orrall, 2024). These measures continue the trend established in 2023, when more than 50 journals were expelled, including prominent titles from publishers such as Hindawi, MDPI, Taylor & Francis, Springer, and Elsevier (Quaderi, 2023). Finally, in the fight against metric fraud in 2025, Clarivate has also announced that citations to retracted articles will no longer count toward the calculation of the impact factor (Arévalo, 2025).

**Analytical functionalities**

Web of Science Core Collection provides an extensive set of integrated analytical tools that enable researchers, institutions, and science evaluators to analyze research output and its impact through multiple perspectives. Within the database itself, we can perform bibliometric analyses through the Citation Reports option, and authors have their own profiles. The JCR, a fundamental tool for this type of analysis and evaluation, offers journal impact metrics including the Journal Impact Factor (JIF), Citation Indicator, Citation Half-life, etc.

Alongside all this, InCites Benchmarking & Analytics[17], its advanced institutional analysis solution, facilitates complex benchmarking from six different analytical perspectives: researchers, research areas, publication sources, organizations, departments, and funding agencies, employing, according to its InCites Indicators Cheat Sheet[18], 88 different indicators for 22 types of entities. The platform integrates tools such as Essential Science Indicators (ESI) to identify relevant research fields. Among recent available improvements, we find Web of Science Research Assistant[19], a tool based on generative artificial intelligence that offers natural language and multilingual search capabilities.

---

[15] https://clarivate.com/academia-government/scientific-and-academic-research/research-discovery-and-referencing/web-of-science/web-of-science-core-collection/editorial-selection-process/
[16] https://clarivate.com/academia-government/scientific-and-academic-research/research-discovery-and-referencing/web-of-science/web-of-science-core-collection/editorial-selection-process/journal-evaluation-process-selection-criteria/
[17] https://clarivate.com/academia-government/scientific-and-academic-research/research-funding-analytics/incites-benchmarking-analytics/
[18] https://incites.zendesk.com/hc/en-gb/article_attachments/24647474424337
[19] https://clarivate.com/academia-government/scientific-and-academic-research/research-discovery-and-referencing/web-of-science/web-of-science-research-assistant/

## 1.1.2. Scopus - Elsevier

**Business model**

Scopus operates under a business model based on institutional subscriptions, managed by Elsevier, one of the largest multinationals in the scientific sector. Access to the platform is commercialized through annual or multi-year licenses aimed at universities, research centers, public organizations, and libraries. The pricing structure is established based on both the research intensity and the size of the subscribing institution, using differentiated scales for consortia and national agreements that may include substantial discounts depending on the type of arrangement. This system contemplates significant annual increases and is subject to periodic negotiations. In this regard, it is a model comparable to that of Web of Science; a model, that of both companies, which many international research centers, including CSIC, encourage to challenge (Ansede, 2024).

**Quantitative coverage**

Scopus stands out for its bibliographic coverage, supported by established relationships with 7,000 publishers from 105 countries, which has enabled it to include 28,791 active peer-reviewed journals (Harling, 2025). In March 2025, they announced that they surpassed the indexing of more than 100 million records[20]. Regarding disciplines, Scopus uses the Science All Journal Classification (ASJC) to classify journals and conference proceedings into four major thematic areas that are further divided into groups and 333 minor fields[21]. One of Scopus's strengths is its extensive open access content, which currently totals 25.1 million articles from 8,137 journals of this type. Additionally, Scopus has expanded its repertoire by including 3.56 million books, distributed across 1,283 collections and 399,000 independent titles, as well as a large amount of content derived from academic activities, with records from 167,000 conference events and 12.9 million presentations[22]. In 2021, preprint indexing was incorporated, which has already accumulated 2.64 million articles from seven specialized servers[23]. Scopus receives more than 95% of data in electronic format from publishers, the content appears indexed on average 2 to 3 weeks after official publication, and articles in press are usually available in less than 4 days after their online version (Elsevier B.V., 2023).

**Editorial criteria**

Scopus applies strict editorial criteria for the selection and maintenance of publications on its platform, under the supervision of the Content Selection and Advisory Board (CSAB), an independent committee composed of recognized experts in various scientific disciplines[24]. Approximately 3,500 inclusion requests are received annually, although only 34% meet the minimum standards to be evaluated, and from this group about 50% is finally accepted after an exhaustive review process by the committee. The selection criteria are structured around five fundamental dimensions (Editorial Policy, Content Quality, Editorial Stability, Publishing Regularity, and Online Availability). Additionally, Scopus requires that articles undergo peer review, preferably under double-blind modality, with the participation of at least two reviewers per issue, as well as compliance with additional mandatory criteria: abstract in English, references in Roman alphabet, valid ISSN, and adherence to ethical standards and good editorial practices aligned with the principles of the Committee on Publication Ethics (COPE). Regarding the expulsion of journals for misconduct, it does not appear to follow a policy as rigorous as Clarivate's, although it is true that there are some documented cases (Travis, 2025).

---

[20] https://blog.scopus.com/100-million-reasons-to-trust-scopus/
[21] https://service.elsevier.com/app/answers/detail/a_id/15181/supporthub/scopus/
[22] https://www.recursoscientificos.fecyt.es/sites/default/files/1%20ScopusIntroducci%C3%B3n.pdf
[23] https://www.elsevier.com/products/scopus/content
[24] https://www.elsevier.com/products/scopus/content/content-selection-and-advisory-board

**Analytical functionalities**

Scopus integrates a broad set of analytical tools. Within the database itself, its functionalities include Analyze Search Results and author profiles that automatically generate a unique identifier to group individual scientific output. It also provides affiliation profiles that allow the evaluation of the research performance of entire institutions. The Journal Analyzer module enables the simultaneous comparison of up to ten journals through graphs of total citations, published articles, and impact trends. Scopus additionally includes three normalized bibliometric indicators that contextualize impact considering disciplinary characteristics. Complementarily, it integrates PlumX Metrics[25], which offers a multidimensional analysis classifying more than 25 metrics into five categories: citations, usage, captures, mentions, and social media. Furthermore, Scopus is closely linked with SciVal[26], a cloud-based analytics platform that uses Scopus data to provide benchmarking tools at the institutional and group level. SciVal allows for the analysis of research trends, identification of key collaborations, and evaluation of scientific impact thanks to its dashboards and 38 indicators distributed across 10 categories[27]. Like Clarivate, they are also working to incorporate and offer AI functionalities[28].

## 1.1.3. OpenAlex – OurResearch

**Business model**

OpenAlex operates as a completely open and free database[29], developed by OurResearch, a nonprofit organization committed to the principles of open science. The complete database dataset is distributed under a CC0 license, allowing any user to access, download, modify, and reuse the data without restrictions or costs. Its financial sustainability is based on a hybrid model with two main revenue sources: (1) premium subscriptions and (2) consulting services. OpenAlex received a $7.5 million grant from Arcadia to establish a sustainable and fully open index of the global research ecosystem over five years[30]. This model contrasts with proprietary databases such as Scopus and Web of Science, which generate revenue through restricted access, and immerse us in a new paradigm of access to scientific Information

**Quantitative coverage**

In a recent literature review on OpenAlex (Forchino & Torres-Salinas, 2025), the reader has access to a comprehensive report on it, but in summary we can say that it has reached an unprecedented scale in the indexing of global scientific literature, housing approximately 260 million bibliographic records by the end of 2024. To achieve this, OpenAlex integrates data from multiple open sources, notably Crossref[31] (150 million works), the legacy of Microsoft Academic Graph, PubMed, arXiv, ORCID, DOAJ, and institutional repositories. The database adds 50,000 records daily and has expanded its primary sources to include DataCite and HAL[32], which has materialized in its recently announced Walden project[33]. Therefore, it includes not only journal articles but also books, chapters, proceedings, datasets, and all types of formats, with data on more than 100 million authors, 110,000 institution headings, and 65,000 Wikidata concepts linked to works[34] (Ho, 2025). To organize scientific knowledge, it uses a 4-level hierarchical classification system based

[25] https://www.elsevier.com/insights/metrics/plumx
[26] https://www.elsevier.com/products/scival
[27] https://elsevier.libguides.com/c.php?g=1328583&p=9781971
[28] https://www.elsevier.com/products/scopus
[29] https://help.openalex.org/hc/en-us/articles/24397762024087-Pricing
[30] https://blog.openalex.org/category/open-science/#:~:text=%247.5M%20grant%20from%20Arcadia
[31] https://help.openalex.org/hc/en-us/articles/24347019383191-Where-do-works-in-OpenAlex-come-from
[32] https://blog.openalex.org/openalex-2024-in-review/
[33] https://blog.openalex.org/openalex-rewrite-walden-launch/
[34] https://docs.openalex.org/api-entities/concepts

on an automated model developed in collaboration with the Centre for Science and Technology Studies (CWTS) at Leiden University that includes 252 subfields[35] and more than 4,000 topics

**Editorial criteria**

OpenAlex adopts a maximum inclusivity policy that contrasts with selective databases such as Web of Science and Scopus. Rather than applying restrictive curation processes, OpenAlex implements comprehensive indexing that integrates data from multiple open sources without prior filters, encompassing journal articles, conference proceedings, books, chapters, datasets, software, and theses. The underlying philosophy maintains that traditional selective curation systematically excludes fields with review standards different from STEM, research in non-English languages, and regional scientific output, so OpenAlex includes all types of journals without analyzing[36] them and transfers the filtering responsibility to users, allowing them to develop lists according to their needs. However, this policy has its consequences.

A brief study based on a sample of 400 journals from the Cabells Predatory Reports[37] catalog revealed that 36.5% of journals identified as predatory are included in OpenAlex, which extrapolated represents approximately between 6,300 and 8,200 predatory journals (Donner, 2025). The most alarming case was that of the publisher OMICS Publishing Group, ordered to pay $50 million by the U.S. Federal Trade Commission, whose articles total more than 215,000 records in OpenAlex. Although OpenAlex offers filters, these were not specifically designed to identify fraudulent journals, so users must be aware of the presence of questionable content in this database.

**Analytical functionalities**

OpenAlex offers analytical capabilities that go beyond basic bibliographic information retrieval. Its REST API[38] provides full-text search of texts, titles, and abstracts with automatic stemming that significantly expands result retrieval, in addition to supporting complex Boolean operators. Users can filter across multiple dimensions. Additionally, OpenAlex provides intuitive web interfaces, bulk data download, access to persistent identifiers (DOI, ORCID, ROR) for interoperability, positioning itself as a versatile platform for creating products and services that require transparent and reusable data.

Leveraging OpenAlex's infrastructure, numerous projects exist that use its open data, including programming libraries such as openalexR and PyAlex, visualization tools such as VOSviewer, analytical dashboards such as the CWTS Dashboard and COKI Dashboard, and search utilities such as SemOpenAlex and oa.m[39].

---

[35] https://help.openalex.org/hc/en-us/articles/24736129405719-Topics
[36] https://help.openalex.org/hc/en-us/articles/27719473439511-How-does-OpenAlex-exclude-predatory-journals
[37] https://cabells.com/solutions/predatory-reports
[38] https://docs.openalex.org/how-to-use-the-api/api-overview
[39] https://help.openalex.org/hc/en-us/articles/27086501974551-Projects-Using-OpenAlex

## 1.2 Selected and recent literature

In this section, we will complement the previous overview through the analysis of the most recent scientific literature on the evaluation of bibliographic databases. We will systematically address five critical dimensions: (1) Total volume of records, analyzing shared corpus and unique content among platforms; (2) Open access journal coverage; (3) Language coverage; (4) Reference coverage; and (5) Metadata quality and reliability, evaluating field availability and accuracy in document classification. This analysis will allow us to identify strengths, limitations, and specific issues of each product.

### 1.2.1 Total volume of records

The analysis of documents indexed by the three databases reveals substantial disparities in their coverage (Table 1): OpenAlex contains 243 million records, compared to 71.2 million in Web of Science and 65.6 million in Scopus (Culbert et al., 2025). When the analysis is delimited exclusively to scholarly articles, OpenAlex maintains its dominance with 200.6 million records, while Web of Science and Scopus present similar figures of 42.6 and 43.5 million respectively. For the specific period 2015-2022 with DOI-based deduplication[40], the figures are: Web of Science 22.6 million records, Scopus 27.6 million, and OpenAlex 76.8 million, representing significant increases compared to previous periods, with Scopus showing greater coverage than Web of Science in the same time Interval.

Table 1. Database coverage according to Culbert et al., 2025

|  | **Web of Science** | **Scopus** | **OpenAlex** |
|---|---|---|---|
| Corpus total | 71,280,830 | 65,642,377 | 243,053,925 |
| Articles | 42,678,632 | 43,579,595 | 200,665,940 |
| 2015-2022 | 22,609,069 | 27,620,472 | 76,836,191 |
| Shared 2015-2022 | 16,788,282 | | |

The analysis of binary intersections between sources reveals asymmetric overlap patterns (Figure 1)[41]. This distribution indicates that no single database captures the entirety of scientific literature. It is significant that Web of Science and Scopus share 90% of their content. On the other hand, the analysis of the "shared corpus"[42] among the three databases for the period 2015-2022 identifies 16.7 million common records with valid DOI, which represents only 74% of Web of Science's corpus, 60% of Scopus's, and barely 21% of OpenAlex's, evidencing the existence of unique content in each platform[43]. This difference reflects the editorial philosophies discussed previously, as Web of Science and Scopus implement a selective model based on editorial curation, while OpenAlex adopts a comprehensive model based on automated harvesting. Another interpretation is that if we make an inverse proportion of uniqueness, it demonstrates that OpenAlex indexes 78.2% of its content uniquely, not shared with Web of Science and Scopus.

---

[40] A deduplication process was followed whereby the DOI is used as a unique identifier to ensure that each publication is counted only once, excluding records without DOI or with duplicate DOIs.

[41] It should be emphasized that Figure 1 underestimates the true overlap, since the DOI-based matching methodology excluded from the analysis 21.7 million OpenAlex records, 4.2 million Web of Science records, and 2.5 million Scopus records that lacked a DOI.

[42] A set of publications indexed simultaneously across all three databases (Web of Science, Scopus, and OpenAlex), identified through DOI matching, enabling direct cross-platform comparisons.

[43] The shared corpus of 16.7 million represents 74% of Web of Science, 60% of Scopus, and only 21% of OpenAlex, meaning that each database contains unique content: 26% in Web of Science, 40% in Scopus, and 79% in OpenAlex. This asymmetry indicates that OpenAlex indexes a significantly larger number of publications not simultaneously available on the other two platforms. In practical terms, relying on a single database means losing access to millions of unique publications indexed in the others.

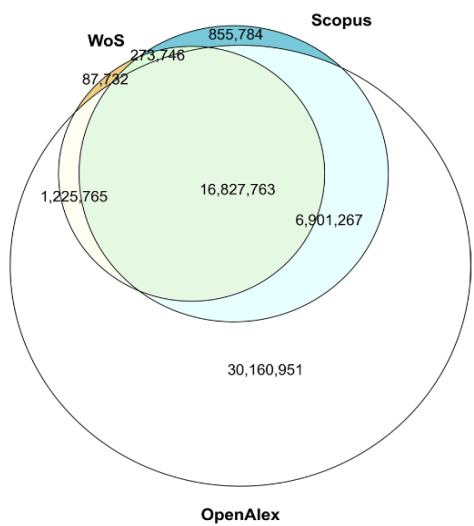

Figure 1. Overlap of different databases according to the study (Culbert et al., 2025).

- 30,160,951: Exclusive OpenAlex records
- 16,827,763: Records shared by all three databases
- 1,225,765: Web of Science records not in Scopus
- 6,901,267: Exclusive Scopus records not in WoS
- 855,784: Records shared only by Scopus
- 273,746: WoS and Scopus records not included in OpenAlex
- 87,122: Exclusive Web of Science records

**Fuente**:
Culbert, J. H., Hobert, A., Jahn, N., Haupka, N., Schmidt, M., Donner, P., & Mayr, P. (2025). Reference coverage analysis of OpenAlex compared to Web of Science and Scopus. Scientometrics, 130(4), 2475-2492

## 1.2.2. Coverage of Open Access Journals

A detailed analysis of the Directory of Open Access Scholarly Resources[44], taking May 2024 as the cut-off date, reveals marked differences in the indexing of open access journals among Web of Science, Scopus, and OpenAlex (Maddi et al., 2025). Out of the 62,701 active open access resources listed in ROAD, Web of Science indexes 6,157 journals (9.8% of the universe), Scopus indexes 7,351 (11.7%), and OpenAlex reaches 34,217 (54.6%), meaning that OpenAlex covers approximately 5.6 times more open access journals than Web of Science and 4.7 times more than Scopus. This difference in volume is also reflected in exclusive coverage (Figure 2): 24,976 open access journals appear only in OpenAlex, compared to just 182 exclusively in Web of Science and 373 only in Scopus, highlighting that OpenAlex indexes a much broader and more diverse universe of open access journals than traditional commercial databases. Finally, only 4,094 open access journals from ROAD are indexed simultaneously in all three databases, i.e., around 6.5% of the total, demonstrating that the actual overlap in coverage is minimal and that each source works with substantially different subsets of the global open access journal ecosystem.

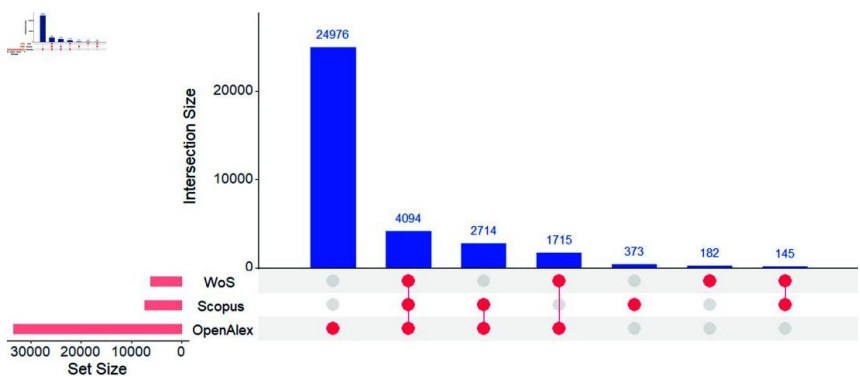

Figure 2. Comparative coverage of open access journals in Web of Science, Scopus, and OpenAlex.

**Source**:
Maddi, A., Maisonobe, M., & Boukacem-Zeghmouri, C. (2025). Geographical and disciplinary coverage of open access journals: OpenAlex, Scopus, and WoS. PLOS ONE, 20(4), e0320347.

---

[44] ROAD (Directory of Open Access Scholarly Resources) is a free service provided by the International ISSN Centre, supported by UNESCO, which offers access to a subset of the ISSN Registry composed of bibliographic records that describe open access scholarly resources (journals, monographic series, conference proceedings, academic repositories, and other serial resources), enriched with metadata drawn from indexing databases, directories, and journal indicators

Combining these results with other studies, we can infer that proprietary databases underrepresent certain geographical regions. A recent study (Khanna et al., 2022) analyzes more than 25,000 journals using the Open Journal Systems (OJS) platform and estimates that they have published around 5.8 million items distributed across 136 countries, with 79.9% of the journals located in the Global South and 84.2% operating under the diamond open access model; despite this, only 1.2% of OJS journals are indexed in Web of Science, while Scopus and OpenAlex index 5.7% and 63.8% of them, respectively. These data, together with those provided by other studies on open access literature coverage (Alonso-Álvarez & Eck, 2025; Simard et al., 2024), help us to understand bibliodiversity in global scholarly communication. All these works show that the dominant scientific evaluation infrastructure perpetuates a structure that systematically marginalizes production from the Global South, where the largest volume of diamond journals is concentrated. This exclusion not only makes invisible the scientific work of entire regions, but also reinforces the hegemony of English and the commercial models of the Global North.

## 1.2.3. Language Coverage

The coverage of languages in major academic databases (Céspedes et al., 2024) reflects a clear predominance of English. In Web of Science and Scopus, more than 95% of indexed articles are published in English, whereas articles in Spanish, Chinese, French, German, Russian, or Japanese reach only values between 0.3% and 1% each. OpenAlex, by contrast, offers a different picture: 68% of its articles are in English, and a notable increase in the representation of other languages is observed, with Chinese (4.2%), Spanish (2.6%), Russian (2.0%), and Japanese (1.3%) standing out, far surpassing the percentages reported for Web of Science and Scopus.

Table 2. Language coverage in the databases according to the work of Céspedes et al., 2024

|  | Web of Science (%) | Scopus (%) | OpenAlex (%) |
|---|---|---|---|
| English | 96.4 | 95.0 | 68.0 |
| Spanish | 0.3 | 0.4 | 2.6 |
| Linguistic diversity | Very low | Very low | Moderate |

## 1.2.4. Reference coverage

An in-depth analysis of the shared corpus of 16.8 million article records published between 2015 and 2022 (see section 1.2.1 and footnote 43) reveals differentiated patterns in bibliographic reference coverage among Web of Science, Scopus and OpenAlex (Cubert et al., 2025). When the comparison is restricted to these overlapping documents, the three databases show comparable averages of source references per article, indicating that, for recent literature indexed simultaneously, they reach similar levels of coverage for source citations. However, the study reveals substantial differences in the ability to capture total references. Web of Science and Scopus register, respectively, an average of 6.4 and 6.2 additional references per article compared to OpenAlex. This difference is concentrated mainly in non-source reference types (grey literature, editorials, theses and other materials without standardized persistent identifiers) and in bibliographic links that could not be normalized[45].

---

[45] OpenAlex's limitation in this respect stems from the fact that its system only counts references classified as sources, which are heavily dependent on the availability of open metadata deposited in repositories such as Crossref, whereas Web of Science and Scopus integrate broader indexing procedures that allow them to include diverse document types and references without fully normalized metadata.

Table 3. Reference coverage according to the work of Culbert et al., 2025

| | Web of Science | Scopus | OpenAlex |
|---|---|---|---|
| Full Database | | | |
| Number of references | 1,765,281,799 | 2,033,522,623 | 1,845,379,285 |
| Avg. nr. of source references per record | 16.867 | 18.692 | 7.572 |
| Shared corpus (2015–2022) | | | |
| Number of references in corpus | 725,008,043 | 727,056,725 | 585,616,069 |
| Avg. nr. of source references per record | 33.416 | 33.363 | 34.863 |

Some case studies with limited samples[46] (Gusenbauer, 2024) based on retrorreference coverage indicators show that Web of Science and Scopus offer a more accurate representation of references. The mean percentage error in prospective counting is very low in Web of Science (+2.9%) and Scopus (+2.8%), implying that both tend to slightly overcount references, whereas OpenAlex presents a pronounced negative bias (−23.9%), i.e., it systematically underrepresents about a quarter of the actual references present in the articles. This difference is also reflected in the mean absolute percentage error, which measures the magnitude of the mismatch: 5.8% in Web of Science and 10.1% in Scopus, compared to 28.9% in OpenAlex. As a result, the composite BWC score, which combines article coverage with reference information and counting accuracy, rates Web of Science at 94.2%, Scopus at 89.9%, and OpenAlex at 68.9%, thus positioning Web of Science as the preferred option for retrospective reference searches, followed by Scopus, and OpenAlex at a considerable distance.

Table 4. Reference coverage indicators according to Gusenbauer (2024)

| Indicators | Web of Science | Scopus | OpenAlex |
|---|---|---|---|
| Mean percentage error (FWC) | +2.9% | +2.8% | −23.9% |
| Mean absolute percentage error (f) | 5.8% | 10.1% | 28.9% |
| BWC score = d·(1−f) | 94.2% | 89.9% | 68.9% |

## 1.2.5. Quality and reliability of metadata

As stated in the book "Bibliometric Sources: A Practical Guide to Selection" (Arroyo-Machado, 2024), the quality and completeness of metadata are key factors in the selection of bibliographic databases. The current landscape shows clearly distinct approaches between established commercial platforms and emerging open-access platforms. On one hand, Web of Science and Scopus have traditionally adopted a selective model that allows them to perform thorough content curation and implement controlled indexing procedures, which results in higher consistency, normalization, and reliability of their bibliographic data. On the other hand, platforms such as OpenAlex prioritize breadth of coverage over fine-grained quality control, favoring scalability and inclusiveness over maximum precision in metadata. This methodological divergence directly affects both the completeness and integrity of the metadata provided and the subsequent data-cleaning and preparation efforts required for bibliometric analysis, ultimately conditioning the validity and reproducibility of bibliometric results. Therefore, it is essential to critically examine the specific characteristics of quality, normalization, and reliability of metadata in each platform.

---

[46] Gusenbauer's (2024) study evaluates 59 citation indexes based on a purposive sample of 259 journal articles selected from nine disciplines in the social sciences, humanities, health sciences, and natural sciences. The sample is drawn from recent systematic reviews and meta-analyses, in order to ensure that documents have sufficiently developed citation networks for comparing the completeness and counting error of citation links across databases. It uses two metrics: forward citation coverage (FWC) and backward citation coverage (BWC).

Document classification reveals notable differences across platforms (Haupka et al., 2024). Web of Science applies a taxonomy with 87 document-type categories, while Scopus uses 18, and OpenAlex evolved from a very limited scheme (3–4 categories) to an expanded typology following revisions introduced between May and July 2024. As can be observed, Web of Science's taxonomy allows for the highest level of granularity, Scopus seeks a balance between specificity and manageability, and until 2024 OpenAlex prioritized simplification. In the pre-May 2024 version, OpenAlex exhibited massive overclassification, labelling 99.42% of the analysed items as "article"; after the update, the distribution better aligns, with articles accounting for about 89.59%. Another study (Mongeon et al., 2025) (Table 5) identifies similar discrepancies in metadata between OpenAlex and Web of Science when analysing more than 6.6 million records with matching DOIs published between 2021 and 2023. The most relevant discrepancies are found in document classification, with over 300,000 cases in which a publication is classified as an article or review in OpenAlex but not in Web of Science, with 93.5% of these representing misclassifications in OpenAlex.

Table 5. Summary of metadata discrepancies detected by Mongeon et al. (2025)

| Metadata | Main feature of the discrepancy |
|---|---|
| Document type | Overclassification in OpenAlex of non-original material as articles/reviews, compared to more specific categories in Web of Science. |
| Language | Divergent assignments of English, with frequent errors in the algorithmic language detection in OpenAlex. |
| Publication Year | Inconsistent use of "online first" year vs. issue year between Web of Science and OpenAlex. |
| Author count | Different treatment of group authors and consortia between Web of Science and OpenAlex. |

Discrepancies are not limited to document classification. Authors (Mongeon et al., 2025) also report differences in publication year, identifying 480,884 discrepancies (8.1% of the total), mainly explained by the fact that OpenAlex tends to record the year of first online publication, while Web of Science records the year of the journal issue. Regarding the number of authors, 71,133 discrepancies (1.2%) were found, largely because both databases handle consortia and research groups listed in the authorship differently. Linguistic discrepancies were less frequent, with 33,516 cases identified (0.6% of the total), although OpenAlex showed a higher error rate (93.9%) in mislabeling non-English articles as English, possibly due to its algorithmic language detection based on title and abstract. Complementary studies confirm that OpenAlex has broader linguistic coverage (75% of articles in English) compared to Web of Science (95% English), but that about 7% of articles are incorrectly classified as English.

The study by Culbert et al. (2025) also provides some interesting information about metadata. For example, covering the period 2015–2022, they identify 4,186,863 records without a DOI in Web of Science, 2,555,909 in Scopus, and 21,709,360 in OpenAlex. The high proportion of documents without identifiers is explained, in the case of Web of Science and Scopus, by the inclusion of document types and sources where the DOI is not mandatory. In OpenAlex, the problem is amplified, probably due to its inherited corpus from Microsoft Academic Graph and its aggregation policies[47]. On the other hand, within the shared corpus (footnote 43), coverage of abstracts shows differences among the three databases. Both Web of Science and Scopus include abstracts in more than 92% of articles, whereas OpenAlex reaches around 87%, i.e., a gap of about five percentage points in abstract availability. Although this quantitative gap is moderate in relative terms, when projected over corpora of tens of millions of records, it implies that OpenAlex lacks abstracts in several hundred thousand articles that do have them[48] in or in Scopus, which limits certain search and content analysis applications.

---

[47] In Web of Science and Scopus, the presence of several million records without a DOI is mainly associated with the inclusion of document types and sources where the use of persistent identifiers is not mandatory or has not been fully implemented (such as certain conference proceedings, books, book chapters, and regional journals), in a context where the DOI infrastructure does not cover disciplines, countries, and languages in a fully homogeneous way. In OpenAlex, this phenomenon is amplified because it draws on a considerably broader corpus inherited from Microsoft Academic Graph, and it also massively aggregates journals, repositories, and other open sources from multiple providers, including many local titles or those with lower editorial standardization, where the DOI has not been assigned or does not appear in the open metadata, which largely explains why it concentrates over 21 million records without a DOI compared to the figures in Web of Science and Scopus.

Although the focus of this study is not on OpenAlex, the issue of metadata is particularly sensitive in this database. In this regard, we refer the reader to our work "The OpenAlex database in review(Forchino & Torres-Salinas, 2025), which we recommend consulting to go deeper into this topic. In that paper, in addition to issues already discussed above (language, publication year, document types, etc.), we report other problems highlighted in the literature. These include, for example, the lack of institutional affiliations: up to 61.5% of OpenAlex records lack any institutional affiliation, with biases that unequally affect certain publishers and regions. We also document evidence that OpenAlex only identifies a fraction of the retractions listed in Retraction Watch, as well as the presence of duplicates, errors in citation assignment, gaps in bibliographic metadata, and a tendency to overestimate the number of references. All of this necessitates implementing additional validation and cleaning routines when OpenAlex is used as a primary source.

Finally, we must note that the above statements, in certain contexts, for specific fields, or based on small, controlled samples, may change (Cebrian et al., 2025). For instance, Spanish authors have shown that, in the health sciences and for a sample of 1,404 articles[49], "no significant differences are observed" between Web of Science, Scopus, and OpenAlex in most of the fields analysed; that is, the three databases provide reasonably complete metadata for the items they index. However, these observations cannot be extrapolated when we refer to the full scope of the databases.

---

[48] Gusenbauer's (2024) study evaluates 59 citation indexes based on a purposive sample of 259 journal articles selected across nine disciplines in the social sciences, humanities, health sciences, and natural sciences. The sample is drawn from recent systematic reviews and meta-analyses, to ensure that documents have sufficiently developed citation networks for comparing the completeness and counting error of citation links across databases. The study uses two metrics: forward citation coverage (FWC) and backward citation coverage (BWC).

[49] The study defines a broad set of essential core fields to assess indexing quality: article title, authors, publication year, source title (journal), volume, issue, starting and ending pages, received citations, DOI, affiliations, abstract, keywords, funding, references, publisher, ISSN, PubMed ID, language, document type, open access status, and authors' ORCID. These fields are extracted for the same sample of 1,404 articles from I3PT (2020–2022) in Web of Science, Scopus, and OpenAlex, and are compared as percentages of the total number of articles effectively indexed in each database, so that "metadata quality" is measured, rather than just coverage level.

# 1.3. Comparative summary table

Table 6. Comparative summary of citation databases: Web of Science, Scopus, and OpenAlex

| FEATURE | WEB OF SCIENCE | SCOPUS | OPENALEX |
|---|---|---|---|
| **BUSINESS MODEL AND ACCESS** | | | |
| TYPE OF ACCESS | Institutional subscription | Institutional subscription | Open and free access |
| OWNER | Clarivate Analytics | Elsevier | OurResearch (non-profit) |
| LICENSE | Proprietary | Proprietary | CC0 |
| **QUANTITATIVE COVERAGE** | | | |
| TOTAL RECORDS | 97 million | 100+ million (Mar 2025) | ~260 million (end 2024) |
| ACTIVE INDEXED JOURNALS | 22,704 journals | 28,791 journals | Data from 150M+ works (Crossref) |
| ARTICLES (FULL CORPUS) | 42.6 million | 43.5 million | 200.6 million |
| ARTICLES (2015–2022) | 22.6 million | 27.6 million | 76.8 million |
| UNIQUE CONTENT | 25.7% unique | 39.2% unique | 78.2% unique |
| SHARED CORPUS (2015–2022) | 74.3% in overlap | 60.8% in overlap | 21.8% in overlap |
| BOOKS | Included in Core Collection | 3.56 million (corpus-wide) | Included in general corpus |
| PREPRINTS | Not directly included | 2.64 million (since 2021) | Included (multiple sources) |
| **OPEN ACCESS COVERAGE** | | | |
| OA JOURNALS IN ROAD | 6,157 (9.8%) | 7,351 (11.7%) | 34,217 (54.6%) |
| EXCLUSIVE OA JOURNALS | 182 unique journals | 373 unique journals | 24,976 unique journals |
| SHARED OA JOURNALS | 4,094 journals (6.5% of ROAD total) | 4,094 journals (6.5% of ROAD total) | 4,094 journals (6.5% of ROAD total) |
| **DISCIPLINARY COVERAGE** | | | |
| MAIN INDEXES / SUBJECTS | SCIE (9,450 journals), SSCI (3,541), AHCI (1,808), ESCI (9,504) | 4 subject areas, 333 ASJC fields | 252 subfields (4-level hierarchy) |
| CLASSIFICATION SYSTEM | 182 SCIE, 58 SSCI, 25 AHCI disciplines | Science ASJC (333 fields) | CWTS-Leiden model (252 subfields) |
| DISCIPLINARY BIAS | Strong STEM (>70%), weak humanities (<5%), social sciences ~15% | Strong STEM (~60%), moderate humanities (~8%), social sciences ~25% | More balanced, better representation of humanities and social sciences |
| **LANGUAGE COVERAGE** | | | |
| ARTICLES IN ENGLISH | 96.4% | 95% | 68% |
| ARTICLES IN SPANISH | 0.3% | 0.4% | 0.026% |
| LINGUISTIC DIVERSITY | Very low | Very low | Moderate |
| **GEOGRAPHICAL COVERAGE** | | | |
| GEOGRAPHICAL BIAS | Strongly Western (35–40% North America, 30–35% Europe) | More Western but less biased than WoS (40% Europe, 30% North America) | More equitable, better representation of the Global South |
| OJS JOURNALS INDEXED | 1.2% | 5.7% | 63.8% |
| **EDITORIAL CRITERIA** | | | |
| SELECTION PROCESS | Very rigorous, 28 criteria (24 quality + 4 impact) | Strict, 5 fundamental dimensions | Maximum inclusivity, no prior filters |
| EVALUATION COMMITTEE | Clarivate subject editors | Content Selection and Advisory Board (CSAB) | Not applicable (automated indexing) |
| ANNUAL REQUESTS EVALUATED | Not specified | ~3,500 submissions (34% meet minimum criteria, 50% of those accepted) | Not applicable |
| PEER REVIEW REQUIRED | Yes, mandatory | Yes, preferably double-blind | Not mandatory |
| DELISTING POLICY | Very strict (17 journals in 2024, 50+ in 2023) | Less strict, documented cases | No curation / delisting |
| PRESENCE OF PREDATORY JOURNALS | Very low (strict control) | Low (moderate control) | 36.5% of a Cabells sample (~6,300–8,200 estimated) |
| **REFERENCE COVERAGE** | | | |
| TOTAL REFERENCES | 1,765 million | 2,033 million | 1,845 million |
| AVG. REFERENCES PER ARTICLE | 16.8 | 18.7 | 7.6 |
| REFERENCES IN SHARED CORPUS | 725 million | 727 million | 585 million |
| AVG. REFERENCES PER ARTICLE ( | 33.4 | 33.4 | 34.9 |
| MEAN PERCENTAGE ERROR (FWC) | +2.9% (slight overcount) | +2.8% (slight overcount) | −23.9% (significant undercount) |
| MEAN ABSOLUTE % ERROR | 5.8% | 10.1% | 28.9% |
| **QUALITY AND RELIABILITY OF METADATA** | | | |
| ABSTRACT AVAILABILITY | >92% | >92% | ~87% |
| RECORDS WITHOUT DOI (2015–2022) | 4.2 million | 2.6 million | 21.7 million |
| DOCUMENT TYPE CATEGORIES | 87 granular categories | 18 categories | 3–4 pre-May 2024, expanded after 2024 |
| DOCUMENT-TYPE CLASSIFICATION ERROR | Very low (~2%) | Low (~1%) | High (~97% of total errors, 93.5% in article/review misclassification) |
| RECORDS WITHOUT INSTITUTIONAL AFFILIATION | Low | Low | High |
| COVERAGE OF RETRACTIONS (RETRACTION WATCH) | High | High | Low (limited fraction identified) |
| DUPLICATE PROBLEMS | Very low | Low | Present |
| ANALYTICAL FEATURES | | | |
| INTEGRATED TOOLS | Citation Reports, JCR, InCites (88 indicators), ESI, Research Assistant (AI) | Analyze Search Results, Journal Analyzer, PlumX, SciVal (38 indicators) | REST API, full-text search, multiple filters, 4,500-topic classification |
| AUTHOR PROFILES | Yes | Yes, with unique identifiers | Yes |
| INSTITUTIONAL PROFILES | Yes (InCites) | Yes | Yes (~110,000 institutional entities) |
| AI TOOLS | Web of Science Research Assistant | In development | Not specified |
| ECOSYSTEM OF EXTERNAL TOOLS | InCites | SciVal | Extensive (openalexR, PyAlex, VOSviewer, CWTS Dashboard, COKI Dashboard, etc.) |
| **UPDATE AND PROCESSING** | | | |
| INDEXING SPEED | Varies by index | ~2–3 weeks on average, in-press <4 days | 50,000 records daily |
| DATA RECEPTION FORMAT | Mixed | >95% electronic format | Multiple open sources |
| TEMPORAL COVERAGE | From 1900–1945 (by discipline) | Mainly from 1996 onwards | Broad temporal coverage |
| **SUSTAINABILITY AND TRANSPARENCY** | | | |
| FUNDING MODEL | Institutional subscriptions | Institutional subscriptions | Arcadia grant ($7.5M), premium subscriptions, consulting |
| TRANSPARENCY | Low (proprietary model) | Low (proprietary model) | High (open source, open data) |
| PROVEN SUSTAINABILITY | Yes (60+ years) | Yes (30 years) | Emerging (from 2022) |
| INTEROPERABILITY | Limited | Limited | High (DOI, ORCID, ROR, Crossref, etc.) |

# PART 2

## Quantitative analysis

In this part of the study, the quantitative analysis with original data is presented; see the methodological annex, covering *Web of Science*, *Scopus* and *OpenAlex* during the period 2015-2024. This section includes the comparison by document types, languages and research areas, allowing observation of the similarities and differences between the three databases and how these variations affect their coverage and representativeness.

## 2.1. Longitudinal analysis

Table 7. Temporal distribution (2015-2024) and by five-year period of journal publications by database

| YEAR↓ | 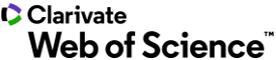 Clarivate Web of Science™ | 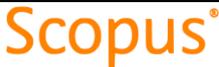 Scopus˙ | 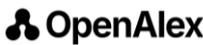 OpenAlex |
|---|---|---|---|
| **2015** | 1,859,511 | 2,157,076 | 5,345,449 |
| **2016** | 1,937,508 | 2,213,298 | 5,314,813 |
| **2017** | 2,008,616 | 2,261,339 | 5,100,536 |
| **2018** | 2,117,402 | 2,377,531 | 5,139,701 |
| **2019** | 2,259,716 | 2,612,050 | 5,361,377 |
| **2020** | 2,491,692 | 2,880,996 | 5,734,366 |
| **2021** | 2,729,213 | 3,110,358 | 5,700,252 |
| **2022** | 2,824,941 | 3,194,670 | 5,228,965 |
| **2023** | 2,736,927 | 3,132,908 | 5,433,988 |
| **2024** | 2,969,954 | 3,389,322 | 5,780,514 |
| **2015-2019** | 10,182,753 | 11,621,294 | 26,261,876 |
| **2020-2024** | 13,752,727 | 15,708,254 | 27,878,085 |
| **TOTAL** | 23,935,480 | 27,329,548 | 54,139,961 |

The temporal distribution of journal publications between 2015 and 2024 (Table 7) shows patterns of sustained growth in the three databases, although with clearly differentiated magnitudes. *Web of Science* increases from 1,859,511 records in 2015 to 2,969,954 in 2024, representing an absolute increase of 1.11 million. *Scopus* evolves from 2,157,076 to 3,389,322, with an increase of 1.23 million, slightly higher than that recorded by *Web of Science*. *OpenAlex* maintains considerably higher figures throughout the entire interval, with values ranging between 5,345,449 in 2015 and 5,780,514 in 2024. The stability of its annual figures contrasts with the lower variability observed between Web of Science and Scopus. In the first five-year period 2015-2019, *Web of Science* accumulates 10,182,753 records compared to 11,621,294 in *Scopus*, whilst *OpenAlex* totals 26,261,876. In the second five-year period 2020-2024, *Web of Science* reaches 13,752,727 publications and *Scopus* 15,708,254, values that represent similar proportional increases between both databases. *OpenAlex* records 27,878,085 documents in this same interval, a figure slightly higher than the previous five-year period. In the cumulative total for 2015-2024, Scopus gathers 27,329,548 documents, surpassing the 23,935,480 of *Web of Science*, whilst *OpenAlex* reaches 54,139,961.

Figure 3 visually represents the same temporal distribution and allows clear observation of the proximity between the trajectories of *Web of Science* and *Scopus* in the period 2015-2024. Both series present progressive and sustained growth that begins at around 1.8-2.1 million documents and exceeds 2.9-3.3 million in 2024. The slope of both curves is similar, and a parallel volume of publications can be appreciated. In contrast, *OpenAlex* is consistently positioned at a considerably higher level with values between five and six million annual records and with oscillations very different from those seen in the other databases. Whilst *Web of Science* and *Scopus* maintain, throughout almost the entire period, constant growth, *OpenAlex* shows reductions in some years, especially in 2022.

Figure 3. Annual distribution (2015-2024) of journal publications by database

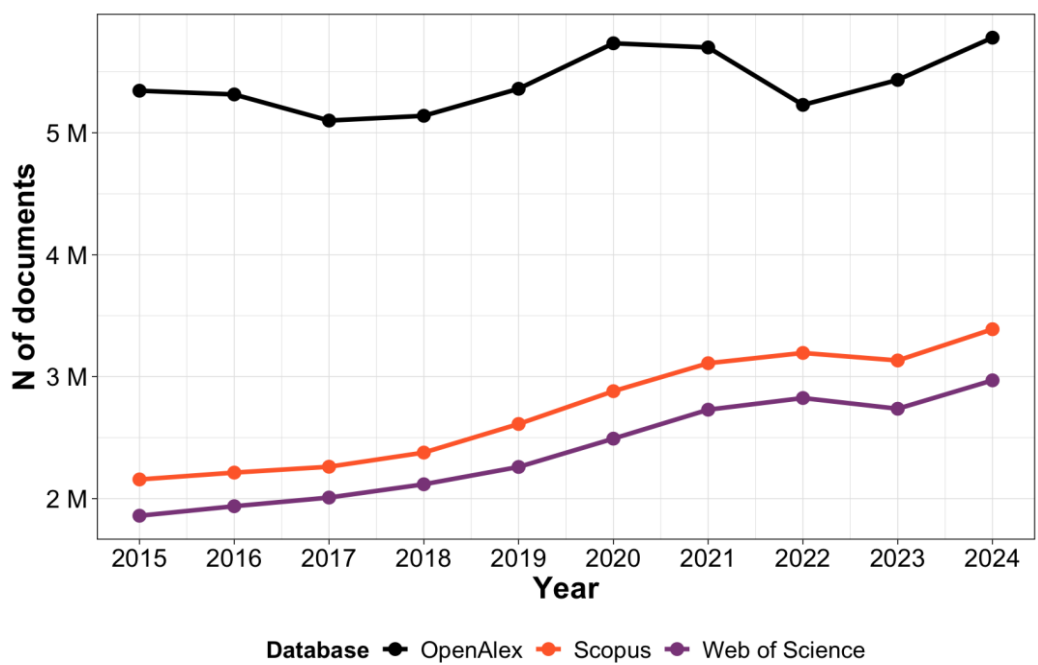

## 2.2. Differences by document types

The distribution by document types shows a very stable pattern in *Web of Science* and *Scopus* (Table 8), where articles concentrate many records throughout the entire period analysed. In the 2015-2024 period, Web *of Science* reaches 21,735,538 articles representing 91% of the total, whilst *Scopus* records 24,683,822 articles equivalent to 90%. The proportion of reviews remains at similar values between both databases, with 7% in *Web of Science* and 8% in *Scopus*. Letters present a reduced and homogeneous weight, around 2% in both cases. This parallelism is also reproduced in the two five-year periods. Between 2015 and 2019, *Web of Science* places articles at 92% and *Scopus* at 90%, whilst between 2020 and 2024 both reach 90% of articles with very close values in reviews and letters. *OpenAlex* presents a different behaviour with higher proportions of articles reaching 95% in the complete period, accompanied by lower percentages in reviews which stand at 3%. The letter category is residual in this database with values close to 1%.

Table 8. Distribution of journal publications by type and database, differentiating by five-year period and total

| TYPE | Clarivate Web of Science™ | Scopus® | OpenAlex |
|---|---|---|---|
| **FULL PERIOD** | | | |
| **Article** | 21,735,538 91% of the total for the period | 24,683,822 90% | 51,654,802 95% |
| **Review** | 1,679,308 7% | 2,126,063 8% | 1,877,982 3% |
| **Letter** | 520,634 2% | 519,663 2% | 607,177 1% |
| **2015-2019** | | | |
| **Article** | 9,321,194 92% of the total for the period | 10,493,146 90% | 25,207,080 96% |
| **Review** | 621,343 6% | 886,970 8% | 742,411 3% |
| **Letter** | 240,216 2% | 241,178 2% | 312,385 1% |
| **2020-2024** | | | |
| **Article** | 12,414,344 90% of the total for the period | 14,190,676 90% | 26,447,722 95% |
| **Review** | 1,057,965 8% | 1,239,093 8% | 1,135,571 4% |
| **Letter** | 280,418 2% | 278,485 2% | 294,792 1% |

Figure 4 offers a visual representation of the distribution by document types for the complete period and allows for clearer appreciation of the differences in magnitude between the three databases. *Web of Science* and *Scopus* shows very close values both in absolute volume and in internal structure. In both, articles occupy the largest part of the total with values close to 22 and 25 million respectively and proportions of around 91% in each case. Reviews are a type of lower volume, around 7-8%, and letters represent an almost residual value, close to 2%. This similarity in distribution is observed both in the absolute numbers graph and in the percentages graph, where the proportions remain practically identical between the two commercial databases. The bar corresponding to *OpenAlex* presents a different behaviour due to its substantially higher volume with more than 51 million articles. The predominance of this category in the internal structure becomes evident in the proportion of 95% that it occupies in the percentages graph.

Figure 4. Absolute and percentage distribution of document types in *Web of Science*, *Scopus* and *OpenAlex* (2015-2024)

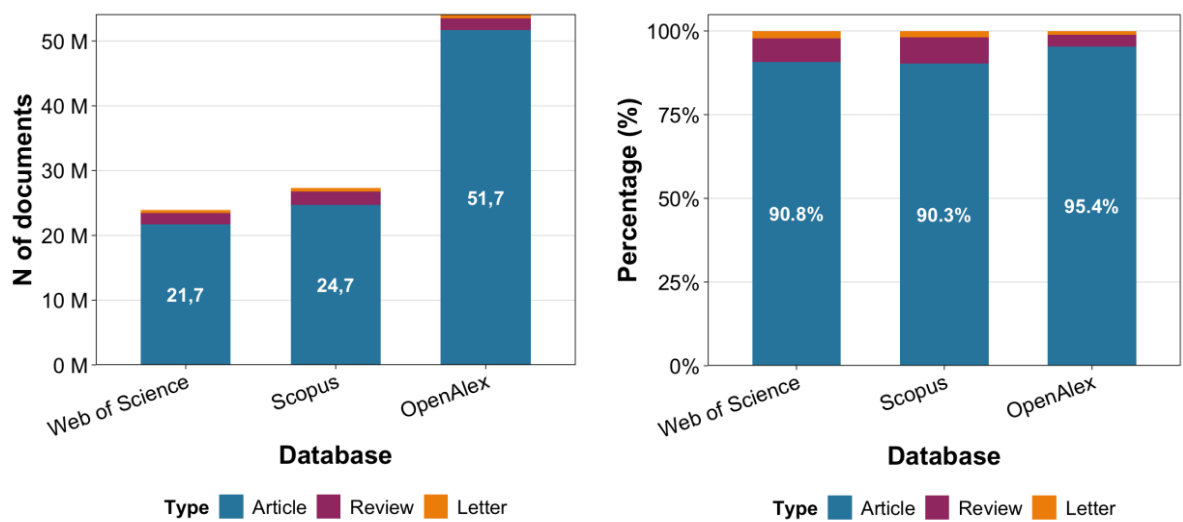

## 2.3. Thematic profiles

The thematic distribution for the complete period (Table 9) shows that Web of Science and Scopus maintain very similar structures in the classification by areas. In both databases, Biomedical and Health Sciences constitutes the main core with 10,178,006 records in Web of Science and 9,932,031 in Scopus, representing 43.3% and 37.3% respectively of the total classified. The areas Physical Sciences and Engineering and Life and Earth Sciences present similar proportions between both databases with values ranging between 14.8 and 23.1% in Web of Science and between 17.7 and 22.5% in Scopus. In Social Sciences and Humanities, the figures vary moderately with 9.0% in Web of Science and 12.1% in Scopus. OpenAlex shows a different distribution with greater weight in Social Sciences and Humanities, which amounts to 7,390,532 records equivalent to 20.4%, and with less predominance of Biomedical and Health Sciences, where it reaches 34.5%[50].

Figure 5 allows visual observation of the thematic composition. In this figure, it is noticeable how in Web of Science and Scopus the production in Biomedical and Health Sciences clearly dominates with proportions exceeding 37% in Scopus and 43% in Web of Science. The remaining areas maintain a balanced distribution with similar presence of Physical Sciences and Engineering and Life and Earth Sciences. Although both databases share a similar thematic pattern, moderate differences are observed in Social Sciences and Humanities, where Web of Science presents 9% and Scopus reaches 12.1%. Likewise, a slight variation is appreciated in Mathematics and Computer Science. The visualisation of OpenAlex presents a different profile with a notable increase in Social Sciences and Humanities, which occupies more than 20% of the total classified, and with a relative weight of Biomedical and Health Sciences lower than that observed in the commercial databases.

---

[50] It is relevant to consider that the thematic classification does not cover the entirety of the corpus due to the algorithmic process employed. *Web of Science* classifies 98% of the total, *Scopus* 97% and *OpenAlex* 67%. This variation in coverage directly affects the comparison given that the unclassified base is substantially larger in OpenAlex, which conditions the relative interpretation of the thematic proportions.

Table 9. Distribution of publications classified by research areas in *Web of Science*, *Scopus* and *OpenAlex* (2015-2024)

| AREA | Clarivate Web of Science™ | Scopus® | OpenAlex |
|---|---|---|---|
| **FULL PERIOD** | | | |
| **Social sciences and humanities** | 2,121,562<br>9.0% of the total classified | 3,229,265<br>12.1% | 7,390,532<br>20.4% |
| **Biomedical and health sciences** | 10,178,006<br>43.3% | 9,932,031<br>37.3% | 12,473,349<br>34.5% |
| **Physical sciences and engineering** | 5,429,077<br>23.1% | 5,984,787<br>22.5% | 6,435,849<br>17.8% |
| **Life and earth sciences** | 3,480,765<br>14.8% | 4,714,799<br>17.7% | 5,556,612<br>15.4% |
| **Mathematics and computer science** | 2,298,690<br>9.8% | 2,784,176<br>10.4% | 4,293,960<br>11.9% |
| **Total** | 23,508,100<br>98% of the total | 26,645,058<br>97% | 36,150,302<br>67% |

Figure 5. Absolute and percentage distribution of research areas classified in *Web of Science*, *Scopus* and *OpenAlex* for the complete period 2015-2024

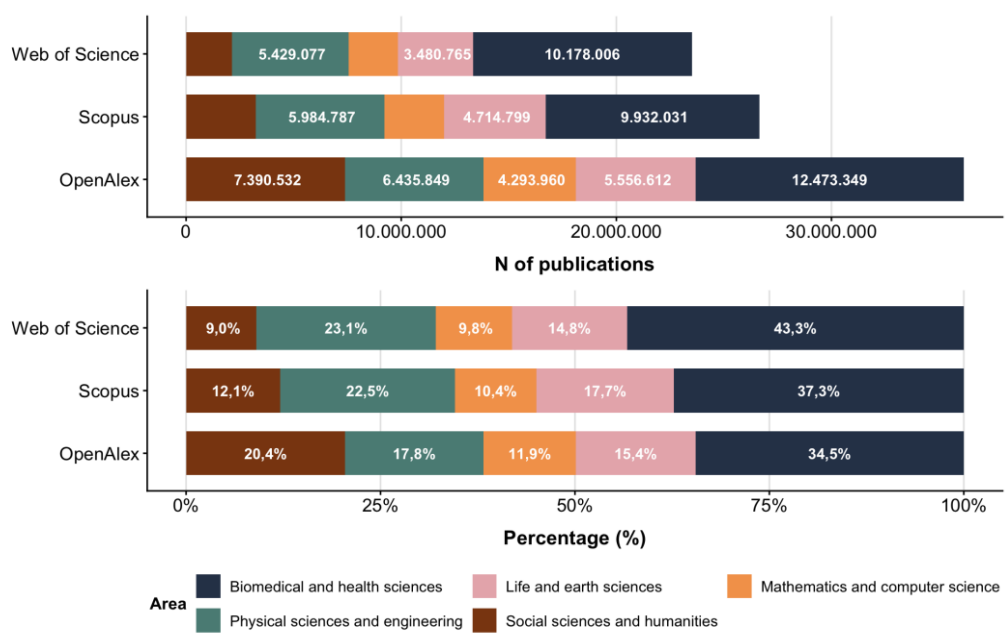

## 2.4. Differences by language

The distribution by language (Table 10) shows a very high concentration in English in the three databases. *Web of Science* records 23,022,125 documents in English equivalent to 96.2% of the total. *Scopus* presents a slightly lower percentage standing at 90.8%. In secondary positions, both databases share similar patterns with reduced presence of Spanish, Chinese, Russian, German and Portuguese, ranging between 0.3% and 1%. These figures highlight a composition with scarce diversity beyond English. The block termed other languages gathers 60 languages in *Web of Science* and 54 in *Scopus*, although with marginal percentages. *OpenAlex* presents a different configuration with 77.5% of documents in English. Spanish occupies the second position with 2.8%, followed by Indonesian with 2.4% and Portuguese with 2.3%, which shows greater linguistic diversity and a more relevant presence of non-Western languages. In the middle part of the ranking, Russian, Japanese, French and German appear with values ranging between 1.5 and 2.2%. The set of other languages in *OpenAlex* reaches 44 languages and totals 2,849,423 documents equivalent to 5.3%, which confirms a more dispersed distribution and a broader capture of multilingual production.

Table 10. Distribution of the main publication languages in *Web of Science*, *Scopus* and *OpenAlex* (2015-2024)

| POSITION | 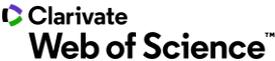 Web of Science | 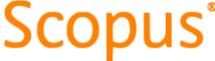 Scopus | 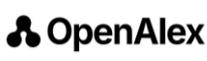 OpenAlex |
|---|---|---|---|
| *1st* | English GB<br>23,022,125 – 96.2% | English GB<br>24,826,757 – 90.8% | English GB<br>41,967,957 – 77.5% |
| *2nd* | Spanish ES<br>262,823 – 1.1% | Chinese CN<br>1,209,576 – 4.4% | Spanish ES<br>1,538,391 – 2.8% |
| *3rd* | Russian RU<br>117,226 – 0.5% | Spanish ES<br>273,381 – 1.0% | Indonesian ID<br>1,316,112 – 2.4% |
| *4th* | German DE<br>103,343 – 0.4% | Russian RU<br>254,843 – 0.9% | Portuguese PT<br>1,241,259 – 2.3% |
| *5th* | Portuguese PT<br>101,569 – 0.4% | German DE<br>158,368 – 0.6% | Russian RU<br>1,177,828 – 2.2% |
| *6th* | Chinese CN<br>87,605 – 0.4% | French FR<br>127,461 – 0.5% | Japanese JP<br>1,077,453 – 2.0% |
| *7th* | French FR<br>83,265 – 0.3% | Portuguese PT<br>91,341 – 0.3% | French FR<br>831,271 – 1.5% |
| *8th* | Italian IT<br>34,716 – 0.1% | Japanese JP<br>76,222 – 0.3% | German DE<br>787,045 – 1.5% |
| *9th* | Turkish TR<br>25,622 – 0.1% | Italian IT<br>51,961 – 0.2% | Korean KR<br>735,882 – 1.4% |
| *10th* | Korean KR<br>18,370 > 0.1% | Korean KR<br>39,255 – 0.1% | Chinese CN<br>615,510 – 1.1% |
| OTHERS | **60 languages**<br>78,816 – 3.3% | **54 languages**<br>220,383 – 8.1% | **44 languages**<br>2,849,423 – 5.3% |

## 2.5. Detail of research areas

Table 11. Detail of research areas according to document type, publications in English and publications with DOI

| | | | Web of Science | | Scopus | | OpenAlex | |
| | | | DOCUMENTOS | % | DOCUMENTOS | % | DOCUMENTOS | % |
|---|---|---|---|---|---|---|---|---|
| **SSH** | | Article | 2,052,532 | 96.7% | 2,962,058 | 91.7% | 7,317,486 | 99.0% |
| | | Review | 62,845 | 3.0% | 249,058 | 7.7% | 69,205 | 0.9% |
| | | Letter | 6,185 | 0.3% | 18,149 | 0.6% | 3,841 | 0.1% |
| | | Pub. in English | 1,934,582 | 91.2% | 2,860,590 | 88.6% | 5,226,194 | 70.7% |
| | | Pub. with DOI | 2,037,193 | 96.0% | 2,967,147 | 91.9% | 6,951,718 | 94.1% |
| **BHS** | | Article | 8,598,877 | 84.5% | 8,328,442 | 83.9% | 10,895,689 | 87.4% |
| | | Review | 1,107,386 | 10.9% | 1,180,386 | 11.9% | 1,282,681 | 10.3% |
| | | Letter | 471,743 | 4.6% | 423,203 | 4.3% | 294,979 | 2.4% |
| | | Pub. in English | 9,754,305 | 95.8% | 9,188,553 | 92.5% | 11,523,340 | 92.4% |
| | | Pub. with DOI | 9,325,621 | 91.6% | 9,533,514 | 96.0% | 12,282,682 | 98.5% |
| **PSE** | | Article | 5,179,302 | 95.4% | 5,696,167 | 95.2% | 6,280,987 | 97.6% |
| | | Review | 243,575 | 4.5% | 278,955 | 4.7% | 151,250 | 2.4% |
| | | Letter | 6,200 | 0.1% | 9,665 | 0.2% | 3,612 | 0.1% |
| | | Pub. in English | 5,354,329 | 98.6% | 5,562,684 | 92.9% | 6,256,346 | 97.2% |
| | | Pub. with DOI | 5,372,293 | 99.0% | 5,840,192 | 97.6% | 6,352,946 | 98.7% |
| **LES** | | Article | 3,275,520 | 94.1% | 4,420,199 | 93.8% | 5,323,138 | 95.8% |
| | | Review | 191,037 | 5.5% | 270,084 | 5.7% | 221,623 | 4.0% |
| | | Letter | 14,208 | 0.4% | 24,516 | 0.5% | 11,851 | 0.2% |
| | | Pub. in English | 3,407,947 | 97.9% | 4,243,497 | 90.0% | 5,185,030 | 93.3% |
| | | Pub. with DOI | 3,395,790 | 97.6% | 4,517,536 | 95.8% | 5,408,595 | 97.3% |
| **MCS** | | Article | 2,239,998 | 97.4% | 2,708,628 | 97.3% | 4,231,447 | 98.5% |
| | | Review | 52,880 | 2.3% | 65,975 | 2.4% | 59,559 | 1.4% |
| | | Letter | 5,812 | 0.3% | 9,573 | 0.3% | 2,954 | 0.1% |
| | | Pub. in English | 2,269,424 | 98.7% | 2,576,189 | 92.5% | 4,037,105 | 94.0% |
| | | Pub. with DOI | 2,253,799 | 98.0% | 2,646,581 | 95.1% | 4,155,320 | 96.8% |

## 2.6. Overlap and convergence of sources

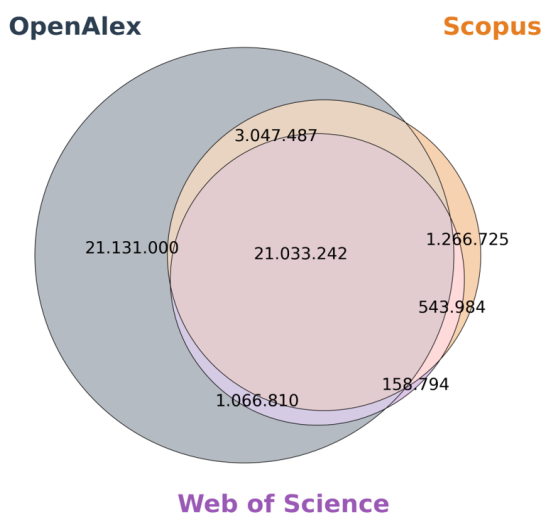

**OpenAlex**  **Scopus**

3.047.487

21.131.000   21.033.242   1.266.725

543.984

1.066.810   158.794

**Web of Science**

Figure 6. Overlap of the different databases according to original data

The comparison between databases shows a very broad common core. As shown in Figure 6 and Table 12, around 21 million documents appear simultaneously in *Web of Science*, *Scopus* and *OpenAlex*, which coincides with the high overlap percentages observed between *Web of Science* and *Scopus* (90% of *Web of Science* in *Scopus* and 79% of *Scopus* in *Web of Science*). *OpenAlex*, although presenting the greatest breadth of coverage, with more than 21 million exclusive records, shows lower integration with the other databases, as only 41% of its documents are in *Web of Science* and 44% in *Scopus*. Even so, it maintains a significant intersection, especially with *Scopus*, whilst *Web of Science* is the database with the lowest volume of exclusive documents (slightly more than one million). Overall, the data emphasise that *Web of Science* and *Scopus* form a highly coincidence and relatively balanced core, whilst *OpenAlex* acts as the most expansive database, combining many its own documents with substantial, but clearly less complete overlap with the other two collections.

Table 12. Pairwise overlap of the different databases according to original data

| Documents from the database in row ↓ present in column → | ◈ Clarivate<br>**Web of Science**™ | **Scopus**˙ | ♣ **OpenAlex** |
|---|---|---|---|
| ◈ Clarivate<br>**Web of Science**™ | | 21,585,087<br>**90%**<br>of WoS in Scopus | 22,100,240<br>**92%**<br>of WoS in OpenAlex |
| **Scopus**˙ | 21,585,087<br>**79%**<br>of Scopus in WoS | | 24,080,992<br>**88%**<br>of Scopus in OpenAlex |
| ♣ **OpenAlex** | 22,100,240<br>**41%**<br>of OpenAlex in WoS | 24,080,992<br>**44%**<br>of OpenAlex in Scopus | |

Table 13 presents, for each of the five scientific areas, Social Sciences and Humanities (SSH), Biomedical and Health Sciences (BHS), Physical Sciences and Engineering (PSE), Life and Earth Sciences (LES) and Mathematics and Computer Science (MCS), the degree of documentary coincidence between *Web of Science* (WoS), *Scopus* and *OpenAlex*. In general, *Web of Science* and *Scopus* show the highest and most stable overlap levels in all areas, with coincidences exceeding 87% and reaching values close to 97% in PSE and LES, indicating very aligned coverage between both databases. In contrast, *OpenAlex* presents greater differences, since, although it gathers significant volumes from each area, its percentage of coincidence with both *Web of Science* and *Scopus* is notably lower, especially in SSH (30%) and in MCS (67%). The most distinctive data demonstrate, on the one hand, the great consistency between *Web of Science* and *Scopus* across all disciplines and, on the other hand, the greater divergence of *OpenAlex*, whose coverage is broad but less overlapping, especially in social sciences and in mathematics and computer science.

Table 13. Pairwise overlap of the different databases considering areas according to original data

| Documents from the database in row ↓ present in column → | 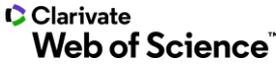 | 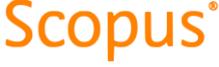 | 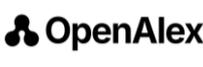 |
|---|---|---|---|
| 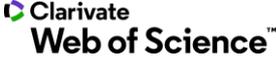 | | **SSH**<br>1,839,398 - **87%**<br>**BHS**<br>8,889,497 - **87%**<br>**PSE**<br>5,269,693 - **97%**<br>**LES**<br>3,295,378 - **95%**<br>**MCS**<br>2,167,372 - **94%**<br>**of WoS in Scopus** | **SSH**<br>1,977,081 - **93%**<br>**BHS**<br>9,070,597 - **89%**<br>**PSE**<br>5,251,121 - **97%**<br>**LES**<br>3,295,987 - **95%**<br>**MCS**<br>2,190,428 - **95%**<br>**of WoS in OpenAlex** |
| 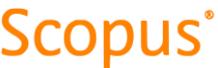 | **SSH**<br>2,228,571 - **69%**<br>**BHS**<br>8,026,133 - **81%**<br>**PSE**<br>5,260,879 - **88%**<br>**LES**<br>3,789,161 - **80%**<br>**MCS**<br>2,180,309 - **78%**<br>**of Scopus in WoS** | | **SSH**<br>2,810,508 - **87%**<br>**BHS**<br>8,971,544 - **90%**<br>**PSE**<br>5,448,461 - **91%**<br>**LES**<br>4,109,580 - **87%**<br>**MCS**<br>2,420,616 - **87%**<br>**of Scopus in OpenAlex** |
| 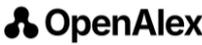 | **SSH**<br>2,183,824 - **30%**<br>**BHS**<br>8,133,579 - **65%**<br>**PSE**<br>5,031,123 - **78%**<br>**LES**<br>3,744,308 - **67%**<br>**MCS**<br>2,633,055 - **61%**<br>**of OpenAlex in WoS** | **SSH**<br>2,598,052 - **35%**<br>**BHS**<br>8,809,335 - **71%**<br>**PSE**<br>5,185,953 - **81%**<br>**LES**<br>4,036,588 - **73%**<br>**MCS**<br>2,856,309 - **67%**<br>**of OpenAlex in Scopus** | |

# PART 3

## 3.1. Executive summary

**Business models and positioning in the ecosystem**
The three databases examined operate under clearly distinct models: Web of Science and Scopus follow long-established subscription-based strategies, with strict curation processes that ensure stability and control. OpenAlex, in contrast, adopts an open, scalable approach, with rapid expansion and dependence on heterogeneous sources. For managers, this poses an operational dilemma: curated reliability versus open breadth. The choice is not mutually exclusive, but it is strategic, and must be aligned with the requirements for evaluation, data auditing, and institutional analysis.

**Coverage and volume differences**
The literature confirms that OpenAlex provides substantially greater coverage, with more than three times as many records as Web of Science or Scopus. This breadth is accompanied by higher rates of unique content and lower overlap with the commercial databases. In contrast, Web of Science and Scopus have similar sizes and a structural overlap of around 90%, which makes them relatively close in real coverage. For managers, volume should not be confused with curated comprehensiveness: each database captures different publication universes, with different underlying scopes and purposes.

**Open Access journals and geographical diversity**
A consistent pattern across the reviewed studies is OpenAlex's greater representativeness in the global Open Access ecosystem, especially for regions and publishers underrepresented in Web of Science and Scopus. This has a direct impact on marginalised disciplines and countries in the Global South whose journals often operate under diamond OA models. Scopus shows intermediate performance, while Web of Science maintains a more selective coverage. This pattern suggests that evaluation decisions must explicitly take into account these structural biases to avoid systematically disadvantaging locally rooted or non-Anglophone research systems.

**Language, references, and citation patterns**
The literature shows that Web of Science and Scopus concentrate over 95% of their content in English, whereas OpenAlex exhibits clearly higher linguistic diversity. In terms of references, Web of Science and Scopus maintain higher levels of coverage and precision, while OpenAlex systematically underrepresents a substantial portion of non-normalised literature, particularly grey material. For citation-based analyses, especially retrospective ones, the commercial platforms still show greater robustness. However, for broad descriptive analyses, OpenAlex provides a more plural and inclusive landscape.

**Metadata quality and operational risks**
Independent studies identify significantly higher rates of errors in document type, publication year, language, and institutional affiliation in OpenAlex, compared to Web of Science and Scopus, whose editorial processes are more consistent and show fewer discrepancies. This does not invalidate OpenAlex's value, but it implies a higher need for prior data cleaning and validation. For managers, the literature is clear: the technical cost of working with open data is greater, but it can be justified when the goal is to broaden the representativeness of analysis beyond the traditional editorial canon.

**Temporal dynamics and growth trends**

Longitudinal analysis shows parallel, stable growth in Web of Science and Scopus between 2015 and 2024, with similar annual increments. OpenAlex maintains a higher overall volume, but with year-to-year fluctuations that reflect changes in source inclusion. For managers, this convergence between Web of Science and Scopus confirms their maturity as consolidated systems, while OpenAlex's fluctuations should be interpreted as part of a more dynamic, expansive ecosystem.

**Document types and structural consistency**

Web of Science and Scopus show virtually identical distributions: around 90% articles, 7–8% reviews, and about 2% letters and other types. OpenAlex, in contrast, raises the proportion of articles to around 95% and drastically reduces other categories. This suggests that the commercial databases offer a more granular document classification, while OpenAlex follows a more simplified scheme. In evaluative contexts, this homogeneity in Web of Science and Scopus facilitates longitudinal comparisons and strengthens the robustness of normalised indicators.

**Topical profiles and disciplinary emphasis**

In topical terms, Web of Science and Scopus display very similar structures, with a clear dominance of Biomedical and Health Sciences and a balanced distribution across other areas. OpenAlex alters this distribution, with a relatively higher share of SSH (Social Sciences and Humanities) and less biomedical weight. This pattern can influence institutional analyses, internal rankings, and diversification studies, as each database reflects a distinct disciplinary landscape that is not fully interchangeable.

**Languages and cultural diversity in scientific production**

Quantitative results confirm that Web of Science and Scopus remain dominated by English-language publications, with over 90% of the total. OpenAlex includes a broader set of languages, expanding the regional and cultural spectrum. For institutions with a significant output in Spanish or non-Anglophone formats, this difference can be decisive. The open database better captures global bibliodiversity, although it requires additional validation due to potential errors in language assignment.

**Global overlap and strategic complementarity**

The final analysis reveals a very extensive common core between Web of Science and Scopus, while OpenAlex contributes the largest volume of exclusive content. The combination of these two realities suggests that no single database, by itself, is sufficient to fully diagnose global scientific activity. For managers, the message is clear: Web of Science and Scopus provide evaluative consistency, while OpenAlex significantly expands representativeness. Their complementary, well-planned use enables a balance between quality control and breadth, thereby reducing disciplinary, linguistic, and geographical biases in bibliometric analysis.

## 3.2. Final recommendations

**Adopt a multi-source strategy with cross-validation.**
Evidence shows that no single database captures the entirety of global scientific production. For rigorous bibliometric analyses, it is recommended to combine sources that balance comprehensiveness (e.g., OpenAlex) with high-quality metadata (e.g., Web of Science, Scopus), using tailored validation protocols according to the specific bibliometric indicators to be calculated. Institutional studies should explicitly report which database is used in each case and provide methodological justification for that choice.

**Define differentiated protocols based on the purpose of the analysis.**
For formal evaluations with implications for research careers, funding, or accreditation, it is advisable to prioritize Web of Science or Scopus, given their higher reliability in critical metadata (affiliations, document types, references). For exploratory analyses, disciplinary mapping, or studies of regional and linguistic coverage, OpenAlex represents a valuable complementary resource. Source selection should carefully balance the required level of comprehensiveness against the acceptable level of metadata precision, depending on the evaluative context.

**Address structural biases in the evaluation of non-Anglophone research.**
Web of Science and Scopus perpetuate systematic underrepresentation of the Global South, non-English languages, and diamond-model publishing. Evaluation processes that rely exclusively on these sources introduce significant distortions against regional research in the humanities, social sciences, and disciplines with local publishing traditions. It is recommended to incorporate assessments of geographical and linguistic bias into evaluative designs, supplementing core databases with regional, specialized sources (e.g., SciELO, Dialnet).

**Include explicit caveats about OpenAlex's limitations.**
While OpenAlex represents a major advance in access to scientific information, its adoption requires clear communication of known operational limitations: documented presence of predatory journals (estimated at 6,300–8,200), systematic errors in document classification, under-counting of references (~24%), gaps in institutional affiliations (61.5% of records lack affiliation), and inconsistent treatment of critical metadata. Reports based on OpenAlex should include specific disclaimers regarding these limitations and, where feasible, cross-validate results through sampling against curated databases.

**Strengthen institutional analytical capacity and continuous monitoring.**
The landscape of bibliographic databases evolves rapidly. It is recommended to invest in sustained training of specialized technical staff and to establish quality protocols for bibliometric analyses that are robust across different data sources. Institutions might also implement periodic monitoring mechanisms that comparatively reassess bibliographic databases as they incorporate technical improvements, update editorial policies, or change coverage, thereby ensuring that strategic decisions are grounded in up-to-date evidence.

# Appendix

## Methods

### 1. Data

The analysis is based on full datasets from Web of Science Core Collection, Scopus, and OpenAlex, downloaded in September, March, and August 2025, respectively. Although the download dates are not fully aligned, this difference does not affect the study, as the analysis is restricted to output indexed up to 2024 and does not consider citations, which are more sensitive to temporal variations. To ensure comparability across platforms, the dataset was restricted to articles, reviews, and letters, applying additional filters as needed. In Web of Science, only journal publications were retained, thereby excluding book chapters and conference proceedings; in OpenAlex, preprints, datasets, and other non-comparable document types were excluded, in line with its inclusive indexing policy.

### 2. Data processing

Internal counts within each database were based on its native unique identifiers: UT (Unique Article Identifier) in Web of Science, EID in Scopus, and Work ID in OpenAlex. These identifiers ensure robust, non-redundant counts within each platform. However, since these identifiers are not interoperable, cross-database comparison was performed using the DOI (Digital Object Identifier) as the linking key, after a normalization process that included corrections and transformations such as case conversion to lowercase. The use of DOIs enabled precise identification of matches, overlaps, and platform-specific content, thereby establishing the comparative structure of the analysis.

### 3. Normalization and thematic analysis

Given that each database uses a different subject classification scheme, the original taxonomies were entirely replaced by applying the Leiden algorithm. This approach allowed the construction of a common hierarchical classification in four levels, ensuring that all three platforms were organized into the same five broad subject areas: Social Sciences and Humanities (SSH), Biomedical and Health Sciences (BHS), Physical Sciences and Engineering (PSE), Life and Earth Sciences (LES), and Mathematics and Computer Science (MCS). This procedure avoids introducing biases related to editorial differences or platform-specific categorization algorithms, and yields a thematic framework that is coherent, reproducible, and fully comparable across databases.

### 4. Tools

Once the data were normalized and the common thematic classification defined, the technical data processing was carried out. Metadata cleaning, deduplication, and standardization were performed using Python 3.13 in JupyterLab 4.3.4, with specific routines for DOI normalization, duplicate control, field validation, and inconsistency detection. Data visualization was generated in R 4.5.1 via RStudio 2025.09.0 Build 387, integrating both environments into a reproducible workflow.

# References

Alonso-Alvarez, P., & Eck, N. J. van. (2025). *Coverage and metadata completeness and accuracy of African research publications in OpenAlex: A comparative analysis* (No. arXiv:2409.01120). arXiv. https://doi.org/10.48550/arXiv.2409.01120

Ansede, M. (2024). El CSIC y medio centenar de organizaciones rechazan las bases de datos privadas que condicionan la ciencia mundial. *El País*. https://elpais.com/ciencia/2024-04-16/el-csic-y-medio-centenar-de-organizaciones-rechazan-las-bases-de-datos-privadas-que-condicionan-la-ciencia-mundial.html

Arévalo, J. A. (2025). Clarivate presenta la edición 2025 del Journal Citation Reports con mejoras en la integridad científica y una cobertura sin precedentes. *Universo Abierto*. https://universoabierto.org/2025/06/19/clarivate-presenta-la-edicion-2025-del-journal-citation-reports-con-mejoras-en-la-integridad-cientifica-y-una-cobertura-sin-precedentes/

Arroyo-Machado, W. (2024). *Fuentes bibliométricas: Guía práctica de selección y uso*. Editorial UOC.

Arroyo-Machado, W., & Torres-Salinas, D. (2025). Incentives accelerate progress on open access in Spain. *Nature*, *646*(8084), 288-288. https://doi.org/10.1038/d41586-025-03276-1

Attride, D., & Orrall, A. (2024). Seventeen journals lose impact factors for suspected citation manipulation. *Retraction Watch*. https://retractionwatch.com/2024/06/27/seventeen-journals-lose-impact-factors-for-suspected-citation-manipulation/

Cebrian, G., Borrego, Á., & Abadal, E. (2025). OpenAlex y Crossref como fuentes de datos bibliográficas alternativas a Web of Science y Scopus en ciencias de la salud. *Revista Española de Documentación Científica*, *48*(1), 1649. https://doi.org/10.3989/redc.2025.1.1649

Céspedes, L., Kozlowski, D., Pradier, C., Sainte-Marie, M. H., Shokida, N. S., Benz, P., Poitras, C., Ninkov, A. B., Ebrahimy, S., Ayeni, P., Filali, S., Li, B., & Larivière, V. (2024). *Evaluating the linguistic coverage of OpenAlex: An assessment of metadata accuracy and completeness* (Versión 2). arXiv. https://doi.org/10.48550/ARXIV.2409.10633

Culbert, J. H., Hobert, A., Jahn, N., Haupka, N., Schmidt, M., Donner, P., & Mayr, P. (2025). Reference coverage analysis of OpenAlex compared to Web of Science and Scopus. *Scientometrics*, *130*(4), 2475-2492. https://doi.org/10.1007/s11192-025-05293-3

Donner, P. (2025). *Prevalence of predatory journals in OpenAlex*. https://www.open-bibliometrics.de/posts/20250603-QuestionableJournals/

Elsevier B.V. (2023). *Scopus content coverage*. Elsevier B.V. https://assets.ctfassets.net/o78em1y1w4i4/EX1iy8VxBeQKf8aN2XzOp/c36f79db2548 4cb38a5972ad9a5472ec/Scopus_ContentCoverage_Guide_WEB.pdf

Forchino, M. V., & Torres-Salinas, D. (2025). *The OpenAlex database in review: Evaluating its applications, capabilities, and limitations*. https://doi.org/10.5281/ZENODO.17357948

Gusenbauer, M. (2024). Beyond Google Scholar, Scopus, and Web of Science: An evaluation of the backward and forward citation coverage of 59 databases' citation indices. *Research Synthesis Methods*, *15*(5), 802-817. https://doi.org/10.1002/jrsm.1729

Harling, L. (2025). Scopus data crosses the 100 million item threshold! *Elsevier Scopus Blog*. https://blog.scopus.com/100-million-reasons-to-trust-scopus/

Haupka, N., Dörner, S., & Jahn, N. (2024). *Recent Changes in Document type classification in OpenAlex compared to Web of Science and Scopus*. https://subugoe.github.io/scholcomm_analytics/posts/openalex_document_types/

Haustein, S., Schares, E., Alperin, J. P., Hare, M., Butler, L.-A., & Schönfelder, N. (2024, julio 23). *Estimating global article processing charges paid to six publishers for open access between 2019 and 2023*. arXiv.Org. https://arxiv.org/abs/2407.16551v1

Hicks, D., Wouters, P., Waltman, L., de Rijcke, S., & Rafols, I. (2015). Bibliometrics: The Leiden Manifesto for research metrics. *Nature*, *520*(7548), 429-431. https://doi.org/10.1038/520429a

Ho, L. H. S. (2025). Can the World's Research Ecosystem be Openly Indexed? *Katina - Librarianship Elevated*. https://katinamagazine.org/content/article/resource-reviews/2024/can-the-worlds-research-ecosystem-be-openly-indexed

Khanna, S., Ball, J., Alperin, J. P., & Willinsky, J. (2022). Recalibrating the scope of scholarly publishing: A modest step in a vast decolonization process. *Quantitative Science Studies*, *3*(4), 912-930. https://doi.org/10.1162/qss_a_00228

Maddi, A., Maisonobe, M., & Boukacem-Zeghmouri, C. (2025). Geographical and disciplinary coverage of open access journals: OpenAlex, Scopus, and WoS. *PLOS ONE*, *20*(4), e0320347. https://doi.org/10.1371/journal.pone.0320347

Mongeon, P., Hare, M., Riddle, P., Wilson, S., Krause, G., Marjoram, R., & Toupin, R. (2025). *Investigating Document Type, Language, Publication Year, and Author Count Discrepancies Between OpenAlex and Web of Science* (No. arXiv:2508.18620). arXiv. https://doi.org/10.48550/arXiv.2508.18620

Quaderi, N. (2023). *Supporting integrity of the scholarly record in Web of Science*. https://clarivate.com/academia-government/blog/supporting-integrity-of-the-scholarly-record-our-commitment-to-curation-and-selectivity-in-the-web-of-science/

Simard, M.-A., Basson, I., Hare, M., Lariviere, V., & Mongeon, P. (2024). *The open access coverage of OpenAlex, Scopus and Web of Science* (No. arXiv:2404.01985). arXiv. https://doi.org/10.48550/arXiv.2404.01985

Torres-Salinas, D. (2024). *Principios de bibliometría evaluativa*. Editorial UOC.

Torres-Salinas, D., Robinson-García, N., & Jiménez-Contreras, E. (2023). La ruta bibliométrica hacia el cambio tecnológico y social: Revisión de problemas y desafíos actuales. *El Profesional de la información*, e320228. https://doi.org/10.3145/epi.2023.mar.28

Travis, K. (2025). Elsevier removes journal from Scopus after Retraction Watch inquiry. *Retraction Watch*. https://retractionwatch.com/2025/05/16/elsevier-removes-journal-from-scopus-after-retraction-watch-inquiry/

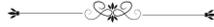

This report was completed in the Nasrid city of Granada on January 22, 2026. This date evokes the memory of María Moliner, Spanish librarian, philologist, and lexicographer, author of the monumental *Diccionario de uso del español*, who passed away the same day in 1981. Her work, developed single-handedly over fifteen years, constitutes one of the most significant contributions to the Spanish language in the twentieth century. During the Second Republic, she carried out intense library work in alignment with the educational modernization and cultural extension projects of her time. After the Spanish Civil War, she became a victim of Francoist repression: purged, barred from official positions, and demoted eighteen ranks in her professional career. Despite this ostracism, she resisted with dignity and channeled her talent into creating her renowned dictionary from her home in Madrid.

¡Viva María Moliner and her legacy!

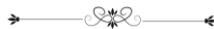

*Influ* Science

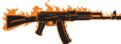

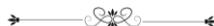